\documentclass[twocolumn,aps,prb,superscriptaddress]{revtex4-2}
\usepackage{mathrsfs}
\usepackage{multirow}
\usepackage{amsmath,amsfonts,amssymb}
\usepackage{array,booktabs}
\usepackage{graphicx,times,graphics,color,epsfig}
\usepackage{extarrows}
\usepackage[english]{babel}
\begin{document}
\title{Giant dynamical electron-magnon coupling in metal-metal-ferromagnetic insulator heterostructure }
\author{Gaoyang Li}
\affiliation{College of Physics and Optoelectric Engineering, Shenzhen University, Guangdong, P. R. China}
%\affiliation{Key Laboratory of Optoelectronic Devices and Systems of Ministry of Education and Guangdong Province, College of Optoelectronic Engineering, Shenzhen University, Shenzhen 518060, China}
\author{Hao Jin}
\email[]{jh@szu.edu.cn}
\affiliation{College of Physics and Optoelectric Engineering, Shenzhen University, Guangdong, P. R. China}
\author{Yadong Wei}
\affiliation{College of Physics and Optoelectric Engineering, Shenzhen University, Guangdong, P. R. China}
\author{Jian Wang}
\email[]{jianwang@hku.hk}
\affiliation{College of Physics and Optoelectric Engineering, Shenzhen University, Guangdong, P. R. China}
\affiliation{Department of Physics, University of Hong Kong, Pokfulam Road, Hong Kong, P. R. China}

\begin{abstract}
Magnon-mediated spin transport across nonmagnetic metal (NM) and ferromagnetic insulator (FI) interface depends critically on electron-magnon coupling. We propose a novel route to enhance electron-magnon coupling dynamically from transport viewpoint. Using non-equilibrium Green's function a theoretical formalism for magnon-mediated spin current is developed. In the language of transport, the effective electron-magnon coupling at NM/FI interface is determined by self-energy of FI lead, which is proportional to density of states (DOS) at NM/FI interface due to nonlinear process of electron-magnon conversion. By modifying interfacial DOS, the spin conductance of 2D and 3D NM/FI systems can be increased by almost three orders of magnitude, setting up a new platform of manipulating dynamical electron-magnon coupling.
\end{abstract}

\maketitle

\section{Introduction}
In conventional spin current studies, the pure spin current without accompanying charge is generated in the nonmagnetic metals (NM). Since this spin current is carried by electron, the waste heat is inevitable. In 2010, it was found that the ferromagnetic insulator (FI) can conduct spin current in the form of magnons without Joule heating and magnons can travel very long distance in YIG\cite{Kajiwara}, which persists even in the presence of disorders\cite{Wesenberg}. Since then, spin transport in FI has become a topic of interest in spintronics.

In the presence of temperature gradient across NM/FI interface, magnon-mediated spin Seebeck effect (SSE) and magnon-mediated spin Peltier effect (SPE) appear. The magnon-mediated SSE was understood in terms of spin pumping and was found to be proportional to the spin-mixing conductance\cite{Xiao}, while SSE was studied in the Pt/YIG bilayer using a linear response theory\cite{Adachi}. Driven by temperature gradient, rectification and negative differential SSE were predicted and rectification of SPE was also discussed\cite{J-Ren}. For a bilayer structure consisting of a paramagnetic metal and FI, the magnon-mediated SPE was studied using non-equilibrium Green's function theory\cite{Ohnuma}. Other studies include noise of spin current\cite{Matsuo} injected by FMR\cite{Kajiwara,Tserkovnyak1}, rectification effect of SSE of a spin Seebeck engine\cite{Tang}, the proposal of optimal heat to spin polarized charge current converter\cite{Sothmann}, the conversion of magnon current to charge and spin current in Coulomb blockade regime\cite{Karwacki}, controlling spin Seebeck current using Coulomb effect in spin Seebeck device\cite{Wu} and magnon-mediated electric current drag in a NM/FI/NM system\cite{Zhang,Han}. The spin Peltier effect in a bilayer structure (paramagnetic metal/FI) and SSE in antiferromagnets and compensated ferrimagnets have also been studied theoretically\cite{Maekawa1,Maekawa2}.

The manifestation of all magnon-mediated spin transport properties studied above depends critically on the magnitude of electron-magnon coupling at the NM/FI interface, which is very small due to the nonlinear process of electron-magnon conversion. In conventional wisdom, the electron-magnon coupling is a static property and the optimal value can be obtained by searching for different materials or interfaces. However, in achieving multi-functionalities, one has to balance among different targeted properties in choosing the suitable material, which makes it difficult to optimize a single property. Efforts have been made to control the quality of interface by avoiding oxidization layer\cite{Nakata}, changing surface roughness\cite{Qiu}, and surface polishing\cite{Aqeel}. Another strategy is to reduce the conductivity mismatch at the interface by inserting another layer of material\cite{Hellman}, which is proven to be very successful. Large enhancement of SSE through NM/FI interface was demonstrated experimentally by inserting atomically thin magnetic and non-magnetic metals, semiconductors, as well as layers of antiferromagnetic insulator (AFI)\cite{Qiu,Aqeel,Lee,Kalappattil,Lee1,Lee2,Wang,Lin}. While reduction of interfacial conductivity mismatch is a general strategy, a new possibility in optimizing effective magnon-electron coupling on top of the static electron-magnon coupling exists. This can be achieved dynamically from the transport point of view by increasing the interfacial density of states (DOS) at NM/FI interface.

It is instructive to recall the electrochemical capacitance where quantum corrections to the classical capacitance can be important in nanoscale systems giving rise to quantum behavior\cite{but1,J-Wang,Hou}. For a parallel plate capacitor, these corrections are determined by the local DOS at the surfaces of two conductors due to the field penetration into the conductors. In contrast to the static DOS, this DOS is for the open system and hence is dynamical\cite{note21}. As a result, the electrochemical capacitance is not purely a geometrical quantity anymore but influenced dynamically by the transport density of states.

It is known that the tunneling structure depicted in Fig.~1(a) can be modeled in Breit-Wigner form. Near the resonance $E=E_0$, the transmission coefficient is given by
\begin{equation}
T = \frac{\Gamma_L \Gamma_R}{(E-E_0)^2 + (\Gamma_L +\Gamma_R)^2/4}
\end{equation}
where $\Gamma_\alpha$ ($\alpha=L,R$) is called the linewidth function characterizing the coupling between the tunneling structure and the $\alpha$ lead with small $\Gamma_\alpha$ corresponding to weak coupling.
%For symmetric system with $\Gamma_L=\Gamma_R$, the transmission coefficient can reach one at resonance with large DOS inside the double barrier.
Thus, in the language of transport, increasing the electron-magnon coupling amounts to increasing the linewidth function at NM/FI interface. Due to the nonlinear process of electron-magnon conversion, the linewidth function of FI lead (the right lead) ${\bar \Gamma}_{R}$ is the energy convolution of DOS matrix at the NM/FI interface and spectral function of FI lead, denoted symbolically as ${\bar \Gamma}_R^0 = D_0 \Gamma_R$ where $D_0$ is the interfacial DOS and $\Gamma_R$ is the spectral function of the FI lead. Therefore, the effective electron-magnon coupling is not a static (geometrical) quantity but is dressed by the transport DOS and can be changed dynamically by incoming electron.
By manipulating the interfacial DOS at NM/FI interface, the effective electron-magnon coupling can be changed drastically. Since the electron-magnon coupling is very small, we model the interface by a barrier with a large barrier height (see Fig.~1(b)). Placing a potential well next to the barrier, electrons are trapped in the well for a long time giving rise to a large interfacing DOS $D_1>D_0$ (Fig.~1(c)). This in turn increases the effective electron-magnon coupling ${\bar \Gamma}_{R}= D_1 \Gamma_R$ making the effective barrier height smaller and thereby increasing the magnon-mediated spin conductance. By putting an additional barrier next to the well (Fig.~1(d)), the electron can resonantly tunnel through the barrier and dwell in the well for even longer time with huge DOS. This yields further increase of effective electron-magnon coupling and reduces the barrier height leading to a huge enhancement of spin conductance. Indeed, in calculating magnon-mediated spin conductance driven by temperature for an ideal 2D (3D) NM/NM/FI nanoribbon (nanowire), a giant enhancement of spin conductance with almost three orders of magnitude is achieved by modifying interfacial DOS. This opens up a new window of engineering the dynamical electron-magnon coupling at NM/FI interface by changing system parameters.

\begin{figure}
\includegraphics[width=8.5cm]{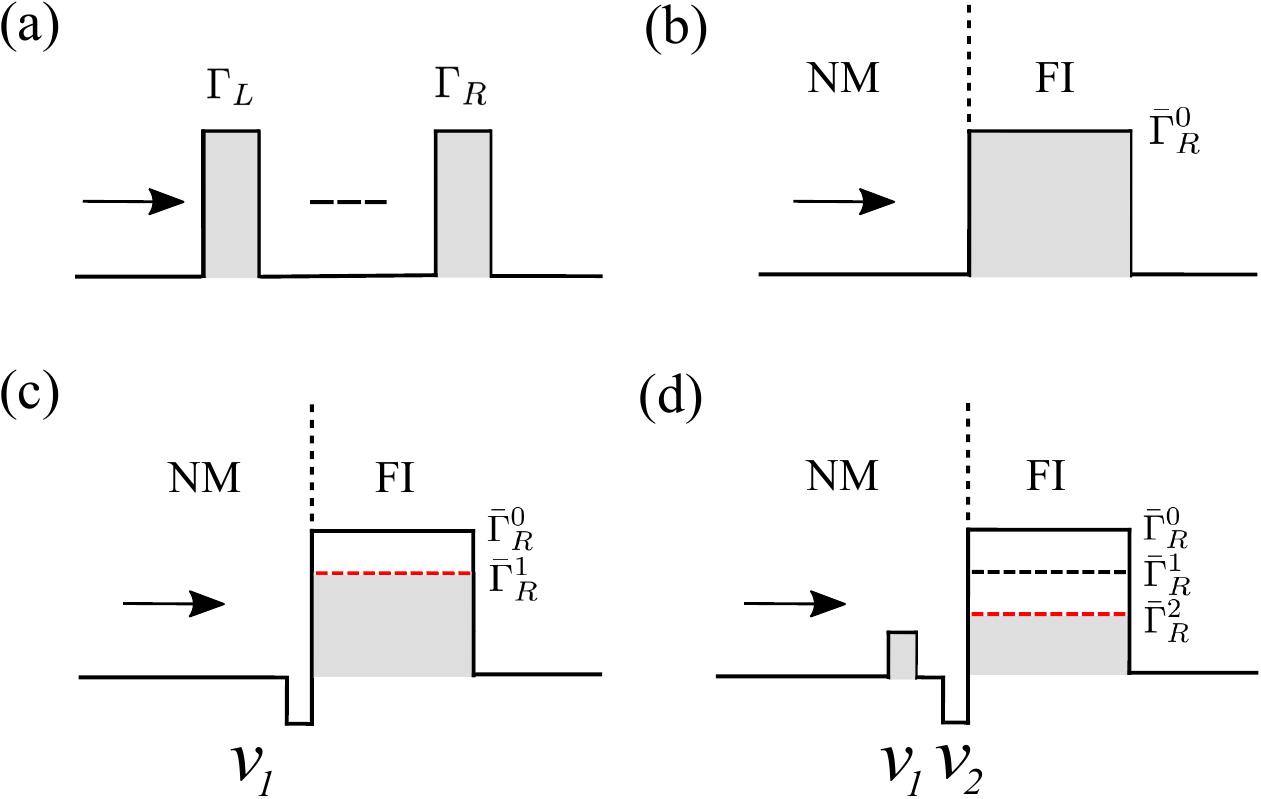}
\caption{(Color online) (a) a resonant tunneling structure where $\Gamma$ describes the coupling between scattering region and the lead. (b) The NM/FI interface is represented schematically by a barrier with an effective coupling constant denoted symbolically by ${\bar \Gamma}_R^0 = D_0 \Gamma_R$. (c). Putting a potential well adjacent to the interfacial barrier gives rise to a large interfacial DOS $D_1>D_0$ and yields a larger effective coupling ${\bar \Gamma}_R^1 = D_1 \Gamma_R$. Note that a larger ${\bar \Gamma}_R$ corresponds to a smaller electron-magnon coupling, the effective barrier height is reduced (denoted by the red line). (d) Placing an additional barrier next to the well generates even larger DOS $D_2$ and effective coupling which corresponds to an even smaller barrier height. }\label{fig1}
\end{figure}

\section{Theoretical formalism}
We consider a system consisting of a central scattering region, a left NM lead and a right FI lead. The Hamiltonian of the system ($\hbar=1$) is given by $H = H_L +H_R +H_d + H_T + H_{sd} $ where $H_L= \sum_{k\sigma} \epsilon_{k\sigma,L} c^{\dagger}_{k\sigma} c_{k\sigma}$ and $H_R= \sum_{q} \omega_q a^{\dagger}_{q} a_{q}$ are Hamiltonians of the left and right lead, respectively. $H_d= \sum_{n\sigma} \epsilon_{n\sigma} d^{\dagger}_{n\sigma} d_{n\sigma}$ is the Hamiltonian of the central scattering region, $H_T=\sum_{k\sigma n} t_{k\sigma n} c^{\dagger}_{k\sigma} d_{n\sigma}+h.c.$ is the coupling between the left NM lead and the central scattering region,
and finally $H_{sd}= -\sum_{q nn'} J_{qnn'} d^{\dagger}_{n\uparrow} d_{n'\downarrow}a^{\dagger}_{q} +h.c.$ is the electron-magnon interaction between the right FI lead and the central scattering region\cite{Zheng,note4}.

\subsection{DC Spin current from the right lead}
The spin current $I_{sR}$ is calculated in Appendix A, we find
\begin{eqnarray}
I_{sR}(t)  =  \int dt'{\rm Tr}[G_{\uparrow}^r(t,t') {\bar \Sigma}_{R\uparrow}^<(t',t)
+G_{\uparrow}^<(t,t') {\bar \Sigma}_{R\uparrow}^a(t',t)]+ h.c.  \nonumber
\end{eqnarray}
where ${\bar \Sigma}^r_{R\sigma} =i G_{{\bar \sigma}}^r .{\tilde \Sigma}_{R\sigma}^< + iG_{{\bar \sigma}}^< .{\tilde \Sigma}_{R\sigma}^a $, ${\bar \Sigma}^<_{R\sigma}=i G^<_{{\bar \sigma}}.{\tilde \Sigma}_{R\sigma}^>$. We note that the electron-magnon self-energy ${\bar \Sigma}_{R\sigma}$ is very similar to the electron-phonon\cite{phonon} or the electron-photon\cite{photon} self-energy which contains non-equilibrium Green's function of the system. The above equation is structurally the same as the current in normal systems except that the self-energies are replaced by the electron-magnon self-energies.

For DC case, we find
\begin{equation}
I_{sR}= \int \frac{dE}{2\pi} {\rm Tr}[G_{\uparrow}^r(E) {\bar \Sigma}_{R\uparrow}^<(E)
+G_{\uparrow}^<(E) {\bar \Sigma}_{R\uparrow}^a(E)+h.c.]
\label{IR}
\end{equation}
where $G^r_{\uparrow}(E)=1/[g_{d\uparrow}^{-1}(E) - \Sigma_{L\uparrow}^r(E)-{\bar \Sigma}_{R\uparrow}^r(E)] = 1/[G_{L\uparrow}^{-1}(E) -{\bar \Sigma}_{R\uparrow}^r(E)] $. After some algebra, Eq.~(\ref{IR}) becomes\cite{note5}
\begin{eqnarray}
I_{sR}&=& -\int dE {\rm Tr}[G_{\uparrow}^r(E) \Gamma_{L\uparrow}(E) G_{\uparrow}^a(E) (i{\bar \Sigma}_{R\uparrow}^<(E)\nonumber \\
&~& + 2f_{L\uparrow}(E){\rm Im}{\bar \Sigma}_{R\uparrow}^a(E))] . \label{isr}
\end{eqnarray}
Since the self-energy ${\bar \Sigma}_{R\uparrow}$ has to be calculated self-consistently for self-consistent Born approximation (SCBA), we first consider BA so that ${\bar \Sigma}^<_{R\uparrow} (t,t')= iG^<_{L\downarrow}(t,t'){\Sigma}_{R}^>(t',t)$\cite{note25}. It is straightforward to show the following relation\cite{note1}
\begin{eqnarray}
{\bar \Sigma}^<_{R\uparrow}(E) = i\int d\omega ~ (1+f_R^B(\omega)) f_{L\downarrow}({\bar E}) D^0_{L\downarrow}({\bar E})\Gamma_R(\omega) \label{sig1}
\end{eqnarray}
where ${\bar E}= E+\omega$, $f_R^B$ is Bose-Einstein distribution for the right lead, $D^0_{L\downarrow}(E)=G_{L\downarrow}^r(E) \Gamma_{L\downarrow}(E) G_{L\downarrow}^a(E)$ is the injectivity of the left lead, a dynamical local DOS matrix for electron coming from the left lead\cite{Buttiker}. Here $\Gamma_{L\downarrow}$ is the linewidth function of the left lead
and $\Gamma_{R}= 2{\rm Im}\Sigma_{R}^a$. Similarly, we find the effective linewidth function of the right lead\cite{note2}
\begin{eqnarray}
{\bar \Gamma}_{R}(E) = \int d\omega ~ (f_R^B(\omega) + f_{L\downarrow} ({\bar E})) D^0_{L\downarrow}({\bar E})\Gamma_R(\omega)
\label{sig2}
\end{eqnarray}
We emphasize that the self-energy ${\bar \Sigma}^{r,<}_R$ and hence $D^0_{L\downarrow}({\bar E})\Gamma_R(\omega)$ are nonzero only at NM/FI interface.
From Eqs.~(\ref{sig1}) and (\ref{sig2}), we arrive at the final result\cite{note7}
\begin{eqnarray}
I_{sR} &&= -\int d\omega (f^B_R(\omega)-f^B_L(\omega))\int dE (f_{L\uparrow}(E)\nonumber \\
&&-f_{L\downarrow}(E+\omega)) {\rm Tr}[A_R(E,\omega)] \label{spin4}
\end{eqnarray}
with
\begin{eqnarray}
A_R(E,\omega)=G_{\uparrow}^r(E) \Gamma_{L\uparrow}(E) G_{\uparrow}^a(E)D^0_{L\downarrow}({\bar E})\Gamma_R(\omega) \label{ar}
\end{eqnarray}
and
\begin{equation}
{\bar \Sigma}^r_{R\uparrow}(E)
%=i \int \frac{d\omega}{2\pi} [G^r_{L\downarrow}({\bar E}) \Sigma_R^<(\omega)
%+  G^<_{L\downarrow}({\bar E}){ \Sigma}^a_{R}(\omega)] \nonumber\\
=\int d\omega [f_R(\omega) G^r_{L\downarrow}({\bar E})
+ i f_{L\downarrow}({\bar E}){\rm Im}G^r_{L\downarrow}({\bar E})]\Gamma_{R} .
\label{sigr}
\end{equation}
%Note that Eqs.(\ref{ar}) and (\ref{sigr}) correspond to BA.
%For SCBA, one has to solve the following two equations self-consistently,
%\begin{eqnarray}
%{\bar \Sigma}^<_{R\uparrow}(E)
%&=& \int d\omega ~ G^r_{\downarrow}({\bar E})[i \Gamma_{L\downarrow}({\bar E}) f_{L\downarrow}({\bar E}) + {\bar \Sigma}^<_{R\uparrow}({\bar E})]\nonumber \\
%&\times& G^a_{\downarrow}({\bar E})(1+f_R^B(\omega)) \Gamma_R(\omega)\nonumber
%\end{eqnarray}
%and
%\begin{eqnarray}
%{\bar \Sigma}^r_{R\uparrow}(E)&=&
%\int d\omega G^r_{\downarrow}({\bar E})[f^B_R(\omega)
%+  (-\Gamma_{L\downarrow}({\bar E}) f_{L\downarrow}({\bar E})\nonumber\\
%&+& i{\bar \Sigma}^<_{R\uparrow}({\bar E}))G^a_{\downarrow}({\bar E})(i/2)]\Gamma_R(\omega)\nonumber
%\end{eqnarray}
%Note that BA has neglected ${\bar \Sigma}^<_{R\uparrow}({\bar E})$ in right hand side of the above two equations and ${\bar \Sigma}^r_R$ in $G^r_{\downarrow}$.

It is easy to see that when spin bias $\mu_s$\cite{note7} and $T_L - T_R$ are all zero, there is no spin current. If we keep only quadratic terms in $J_{qn n'}$ and treat Green's functions as scalars (zero dimensional system), Eq.~(\ref{spin4}) recovers the spin current found in Ref.~\onlinecite{J-Ren}, Ref.~\onlinecite{Tang}, and Ref.~\onlinecite{Wu} which are for zero dimension only and were obtained from the equation of motion, full counting statistics formalism, and nonequilibrium Green's function theory, respectively.

\subsection{Electron-magnon conversion}
From Eq.~(\ref{isr}), it is easy to see that if the right lead were replaced by a normal lead, then $\Sigma^<_R = f_R (\Sigma^a_R -\Sigma^r_R)$ and $\Sigma^a_R -\Sigma^r_R = i\Gamma_R$,  Eq.~(\ref{isr}) would formally resemble the usual Landauer Buttiker formula. In realistic systems, the electron-magnon coupling is usually very small\cite{Zheng}, nevertheless, it can be modified dynamically. From Eq.~(\ref{sig2}), the effective self-energy of the FI lead is given by $D^0_{L\downarrow}({\bar E})\Gamma_R(\omega)$ which is proportional to the dynamical DOS at the NM/FI interface. Hence the effective self-energy of the FI lead can be increased by increasing local DOS at the interface. Since the electron-magnon coupling can be changed dynamically during transport, we term it as dynamical electron-magnon coupling\cite{note22}.
%As will be demonstrated by model calculation below, simple modifications of interfacial DOS can drastically increase dynamical electron-magnon coupling in 2D systems.

\section{Numerical results}
In the linear response regime, we define spin conductance driven by temperature gradient as $I_{s} = G_T \Delta T$.
To perform numerical calculation on spin conductance, we have to determine the self-energy ${\bar \Sigma^{r,<}_R}$ defined in Eqs.~(\ref{sig1}) and (\ref{sigr}).
%Since it is difficult to determine the matrix element of $J_{qn n'}$, approximation has to be made. In the tight binding (TB) representation, indices $n,n'$ in $J_{qn n'}$ can be treated as lattice labels. In this representation, the coupling between FI lead and NM region is nonzero only at the interface. One simple approximation is
Assuming that $J_{qn n'} = J_{q} \delta_{n n'}\delta_{n i}$ which corresponds to destroying a magnon and creating two electrons with opposite spins at the same lattice site $i$ at the interface,
%Under this approximation,
we have
\begin{eqnarray}
{\bar \Sigma}_{R\uparrow n_1n}(\tau,\tau') = i\Gamma_0 G_{ n_1 n\uparrow}(\tau,\tau') \Gamma_0 \Sigma_{R}(\tau',\tau)\label{self-appr}
\end{eqnarray}
where  $\Gamma_0$ is a diagonal matrix with nonzero matrix elements at the interface of right lead and $ \Sigma_{R}(\tau,\tau') = \sum_{q} J^2_{q} g_{Rq}(\tau,\tau')$.

We further assume that the spectral function of the magnonic reservoir is Ohmic so that\cite{J-Ren,Tang,Wu,Book}
$\Gamma_R(\omega)=\alpha \omega e^{-\omega/\omega_c} t \Gamma_0$,
where $\alpha$, which is proportional to $J_q^2$, is the dimensionless effective coupling energy between NM and FI and $t$ is the hopping constant. As discussed in Ref.~\onlinecite{{Zheng}}, $\alpha$ is related to spin mixing conductance\cite{Tserkovnyak1,Brataas}.

%Under these approximations, Eqs.(\ref{ar}) and (\ref{sigr}) can be written as
%\begin{eqnarray}
%A_R(E,\omega)=\Gamma_0 D_{L\uparrow}(E) \Gamma_0 D^0_{L\downarrow}({\bar E})\Gamma_R(\omega) , \label{ar1}
%\end{eqnarray}
%\begin{eqnarray}
%{\bar \Sigma}^r_{R\uparrow}(E)
%=\int d\omega \Gamma_0[f_R(\omega) G^r_{L\downarrow}({\bar E})
%+ i f_{L\downarrow}({\bar E}){\rm Im}G^r_{L\downarrow}({\bar E})]\Gamma_{R} . \nonumber
%%\label{sigr1}
%\end{eqnarray}
%Similar expressions can be written down for SCBA. From Eq.(\ref{ar1}) we see clearly that the matrix $A_R$ and hence the spin conductance is determined by dynamical DOS matrix $D^0_{L\downarrow}$ at the interface.

In the numerical calculation, we use the tight-binding approach to discretize the 2D and 3D systems\cite{Datta} and the self-energy of the left lead is calculated using the transfer matrix method\cite{self}, from which the Green's function $G^r_{L\downarrow}$ can be calculated. Once $G^r_{L\downarrow}$ is obtained ${\bar \Sigma}^r_{R\uparrow}$ and $G^r_{\uparrow}$ can also be calculated. Finally, one has to perform a double integration over energy and frequency even for spin conductance, which is very time consuming.

\subsection{Enhancement of spin conductance}
We consider a 2D system using $20\times 20$ mesh with lattice spacing $a=5nm$ so that the hopping parameter $t = 21.8meV$. We consider the ballistic regime and set potential of NM regime to be a constant. We fix the temperature $T=5K$, $\omega_c = 0.24meV$, and $\mu_L = 0.8meV$ which corresponds to the first subband transport. As will be seen below, the enhancement effect mainly comes from the first subband transport.
\begin{figure}
\includegraphics[width=8.5cm ]{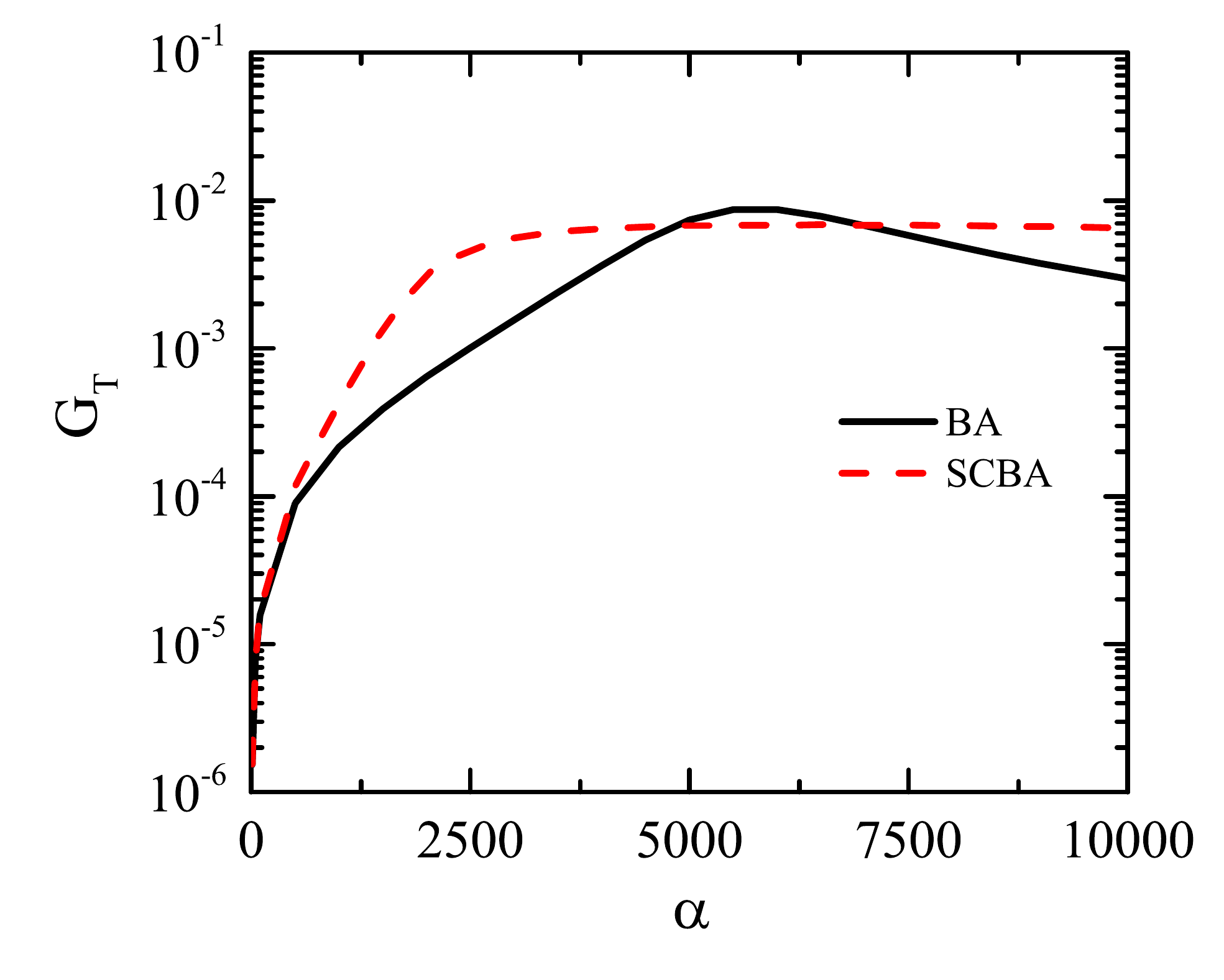}
\caption{(Color online) 2D spin conductance vs $\alpha$ for the clean system.}\label{figsm2}
\end{figure}

%\noindent{\bf Enhancement of 2D spin current as a function of parameter $\alpha$}

Since we consider the ballistic regime, the material ingredients are manifested in the electron-magnon coupling constant. In Fig.~2, the 2D spin conductance vs $\alpha$ is depicted for the clean system ($v=0$) using BA and SCBA showing non-monotonic dependence. The maximum spin conductance is about $7\times 10^{-3}$ at $\alpha>2500$ in SCBA setting up an upper bound for the chosen system parameters. We here choose the most popular (optimal) NM/FI interface, Pt/YIG, as an example with a suitable parameter $\alpha$ ($\alpha=[100,1000]$) that is within the estimated range of electron-magnon coupling for that interface. We choose $\alpha=100$ such that the self-energy ${\bar \Sigma}^r_R$ is in the order of magnon energy (less than $kT$), which is equivalent to $\eta=1$ where $\eta$ is the effective coupling constant at the Pt-YIG interface defined in Ref.~\onlinecite{Zheng}.

To change the interfacial DOS, we modify the potential landscape of three layers next to the interface. Fig.~4(a) shows the schematic plot of the NM/NM/FI system where we label three layers of sites near NM/FI interface with potential $v_1$, $v_2$, and $v_3$, respectively. We consider four typical configurations (from $c_1$ to $c_{4}$): $(v_3,v_2,v_1)= (0,0,-v)$; $(0,v,-v)$; $(-v,0,-v)$; $(v,0,-v)$. They can be classified into three categories: one-layer configuration $c_1$; two-layer configuration $c_2$ and three-layer configurations ($c_3$ and $c_{4}$).
%In Fig.1a, spin conductance vs $\alpha$ is depicted for the clean system ($v=0$) using BA and SCBA showing non-monotonic dependence. The maximum spin conductance is about 0.3 at $\alpha=6\times 10^{3}$ in SCBA setting up an upper bound for the chosen system parameters. From now on, we set $\alpha=600$.
For the clean system, SCBA gives $G_T = 1.7 \times 10^{-5}$ (pink star in Fig.~3) which is the reference for the enhancement of spin conductance.

\subsection{One-layer strategy}

\begin{figure}
\includegraphics[width=8.5cm ]{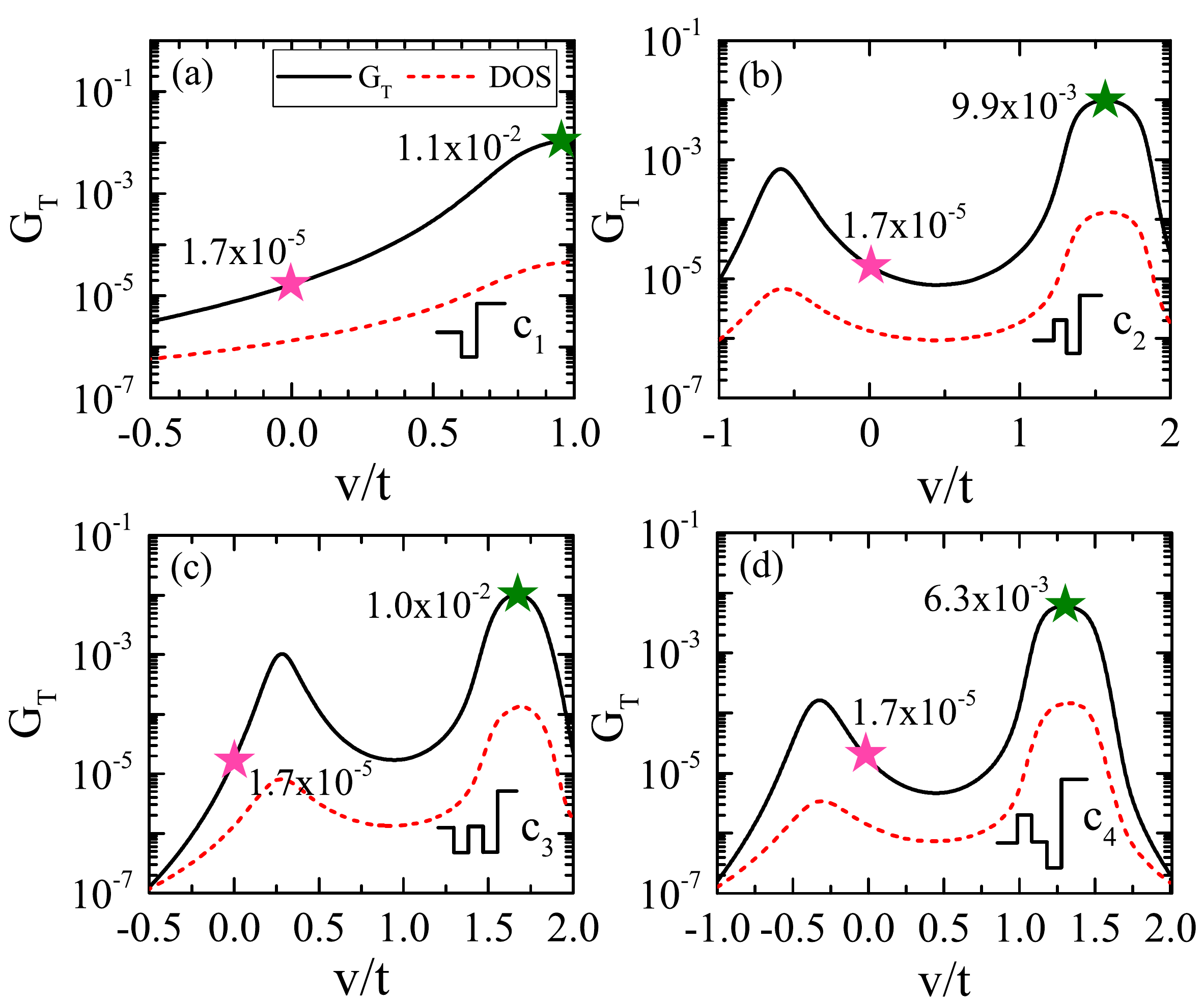}
\caption{(Color online) 2D spin conductance (solid line) and the corresponding interfacial DOS (dashed line) vs $v$ for different configurations $c_1$ to $c_4$ (in panels (a) to (d)) in SCBA at $\alpha=100$. The pink star is the reference with no interfacial modification while the green star shows the largest enhancement. Insets: potential profiles for the corresponding configurations.}\label{fig2}
\end{figure}

One-layer configuration corresponds to insertion of one "layer" of intermediate material at the Pt-YIG interface. In Fig.~3(a), spin conductance and the corresponding interfacial DOS vs $v$ are shown for this configuration in SCBA. We see that the spin conductance and the corresponding interfacial DOS are well correlated showing the increasing of interfacial DOS is the microscopic mechanism for the enhancement of spin conductance. In addition, spin conductance enhancement occurs when the intermediate layer has a lower work function than that of Pt (corresponding to positive $v$). For configuration $c_1$, the largest enhancement can reach 647 times (comparison between the green star and pink star in Fig.~3(a)). On the other hand, inserting an intermediate material with a higher work function (corresponding to negative $v$) suppresses the spin conductance. Our result is consistent with experimental results that enhancement between $300\%$ to $600\%$ were achieved when intermediate materials such as Ru\cite{Nakata}, monolayer WSe$_2$\cite{Lee}, multilayer MoS$_2$\cite{Lee1,Lee2}, and C$_{60}$\cite{Kalappattil} are inserted at the Pt-YIG interface\cite{note10}. The enhancement has previously been attributed to the reduction of conductivity mismatch at the Pt/YIG interface, i.e., the spin mixing conductance $g_{\rm Pt/YIG}$ is smaller than the total conductance of Pt/X/YIG trilayer\cite{note13}. Since interfacial conductance has been considered implicitly in potential profile of the TB model, we interpret it as due to the enhancement of the interfacial DOS\cite{note11}. Moreover, when a nano-scale amorphous layer is formed at the interface\cite{Qiu} or the interface is oxidized\cite{Nakata}, the spin conductance is suppressed\cite{note14} which is also consistent with our result.

\begin{figure}[b]
\includegraphics[width=8.5cm ]{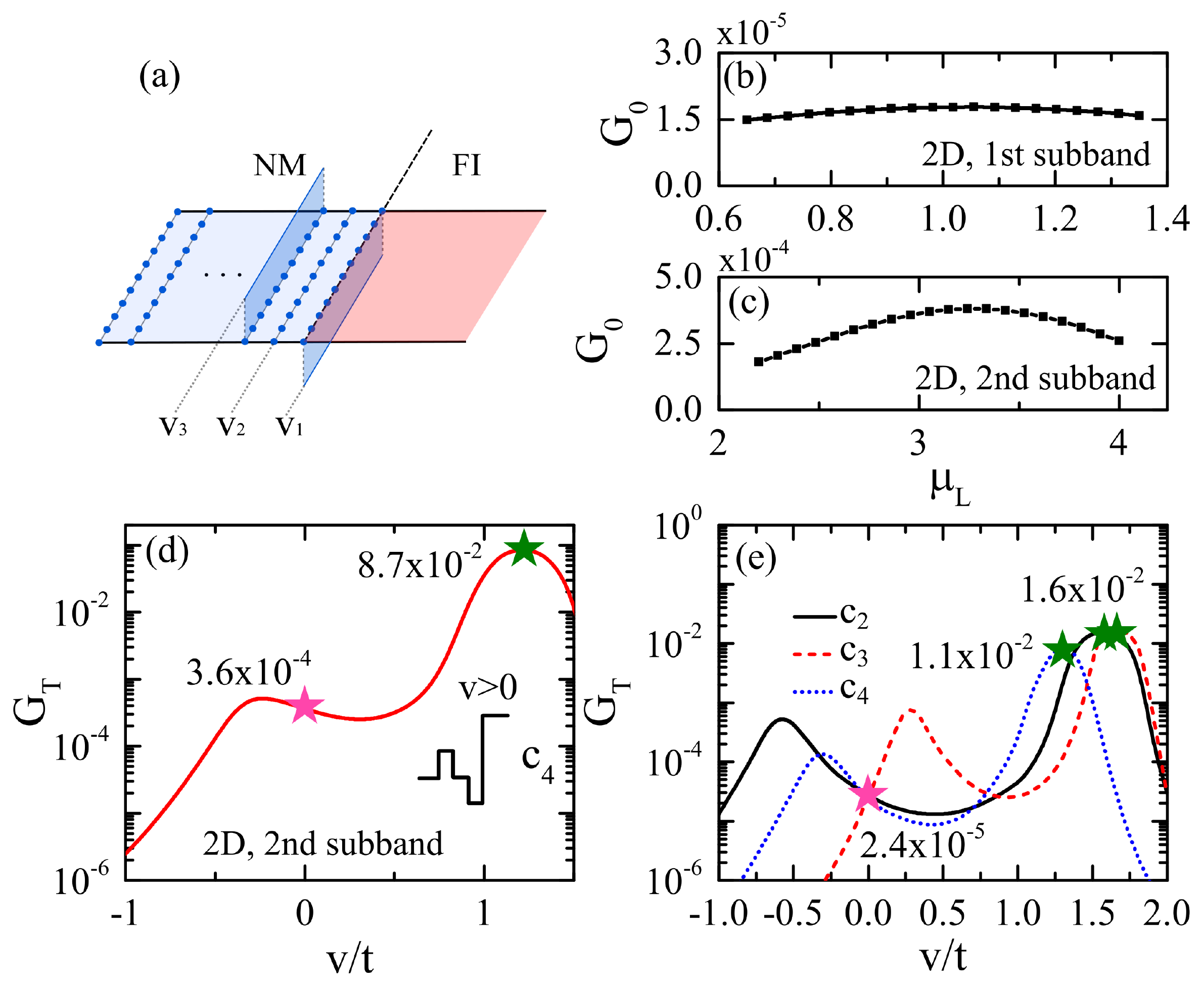}
\caption{(Color online) (a) Modification of interfacial DOS by insertion of different materials at different layers. Three layers with potential height $v_i$ are illustrated. (b) and (c) 2D spin conductances of clean system vs Fermi energy (in unit of meV) in the first and second subband, respectively. (d) 2D spin conductance vs $v$ for configuration $c_4$ when incident electron is in the second subband, yielding a maximum of 242 enhancement. (e) 3D spin conductance vs $v$ for configurations $c_2$ to $c_4$ when incident electron is in the first subband. The enhancement is about 708, 708, and 416 times, respectively for $c_2$ to $c_4$. Here the system is a nanowire with cross-section $20 \times 20$ and $\alpha=100$ in 3D.}\label{fig3}
\end{figure}

\subsection{Two-layer strategy}
Fig.~3(b) shows spin conductance and interfacial DOS as a function of $v$ for "two-layer" configuration $c_2$ where two different materials are needed to insert at the interface. When $v$ is negative, a factor of 44 enhancement can be obtained. When $v$ is positive, a potential barrier is created followed by a potential well near the interface (see inset of Fig.~3(b)) so that an incident electron can dwell for a long time inside the well giving rising to a very large DOS at the interface, which in turn leads to a spin conductance enhancement of 582 times. It is clear that adding a potential barrier increases interfacial conductivity mismatch. Since the conductance mismatch at NM/FI interface can be viewed as a "potential barrier" at the interface, introducing a second barrier in the configuration $c_2$ demonstrates the importance of the "double barrier" structure which has not been explored experimentally.

\subsection{Three-layer strategy}
Fig.~3(c) and Fig.~3(d) depict spin conductance and interfacial DOS vs $v$ for the configurations $c_3$ to $c_{4}$. The configuration $c_3$ is a double well structure and an enhancement of 588 times can be achieved. The configuration $c_4$ corresponds to "double barrier" structure similar to $c_2$. As expected from the result of $c_2$, a large enhancement of 371 times is obtained.

Note that the largest enhancement occurring at a particular potential parameter in $c_1$ to $c_4$ is a typical resonant behavior. If an incident electron has two transmission channels, the longitudinal energies of each channel are different\cite{note26}. Therefore, only one of the transmission channels can reach resonance. As a result, the largest enhancement is expected to reduce by a factor of two which is confirmed numerically (reduced from 371 times to 242 times, see Fig.~4(d) compared to Fig.~3(d)). This shows that the giant enhancement of spin conductance is prominent only in the first subband transport. A giant enhancement of spin conductance with 708 times in 3D nano-wire NM/NM/FI system is also achieved when incoming electron is in the first subband (see Fig.~4(e)).

We point out that our model calculation is carried out in a 2D nanoribbon and 3D nanowire within TB approach and may not be applied directly to realistic systems. In addition, the enhancement effect is prominent only in the first subband. Nevertheless, our theory provides a physical understanding of enhancement of magnon-mediated spin conductance and a general theoretical guidance for interface engineering of dynamical electron-magnon coupling. Although large enhancement of spin conductance is obtained theoretically in this work, it is still a challenging task for experiments.

In this work, besides Born approximation we also make approximation on static electron-magnon coupling, neglected the dipole-dipole interaction in FI and possible dephasing mechanism. We note that the enhancement of dynamical electron-magnon coupling is on top of the static electron-magnon coupling so the approximation made on static electron-magnon coupling does not matter. In addition, the enhancement is due to the increasing of local DOS in the neighborhood of the NM/FI interface where dephasing can be neglected. Furthermore, as shown in Fig.~5, enhancement of 760 to 1000 times are achieved for configuration $c_4$. The large enhancement of 2D spin conductance does not depend on Fermi energies and temperatures suggesting that it is a generic feature.

\begin{figure}
\includegraphics[width=8.5cm ]{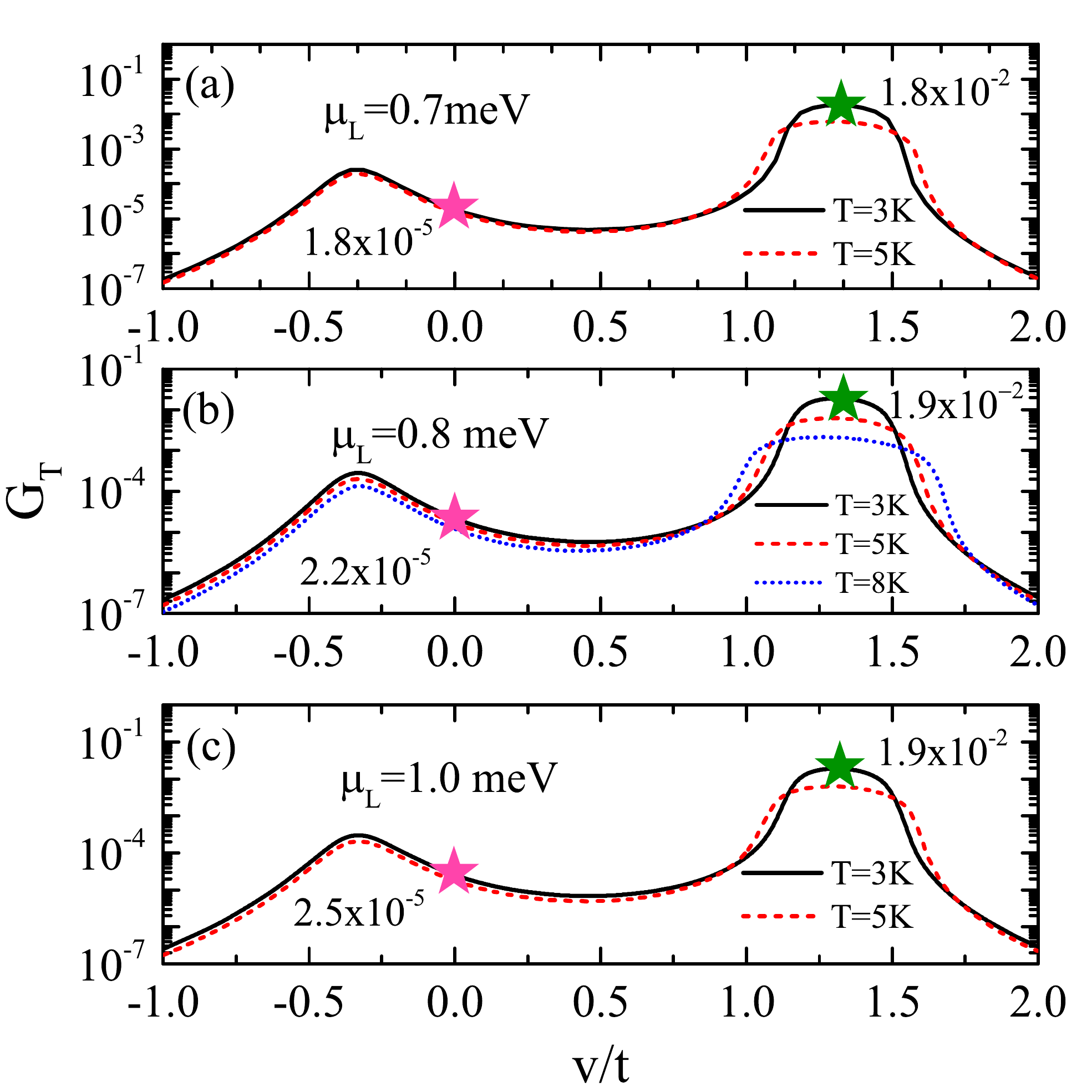}
\caption{(Color online) 2D spin conductance vs $v$ for configuration $c_4$ at different temperatures and Fermi energies at $\alpha=100$.}\label{figsm4}
\end{figure}

\section{Conclusion}
We have developed a theoretical formalism based on non-equilibrium Green's function method for magnon-mediated spin current in NM/NM/FI heterostructure. Because the conversion of electron-magnon across NM/FI interface is a nonlinear process, the effective electron-magnon coupling of NM/FI interface is an energy convolution of local DOS of the system with the spectral function of the FI lead, in distinct contrast to normal systems. Given that the electron-magnon coupling is very small, this provides a new possibility to manipulate the coupling dynamically by modifying potential landscape near the interface. As demonstrated in this work, simple modification of interfacial DOS can drastically increase the spin conductance in 2D and 3D systems as a result of enhancement of dynamical electron-magnon coupling.

\section*{ACKNOWLEDGEMENT}
This work was financially supported by the Natural Science Foundation of China (Grant No. 12034014).

%\clearpage
\appendix

\section{Calculation of Spin Current}

We partition $H$ into $H_0$, Hamiltonian of the isolated leads and the central scattering region, and coupling terms $H'$. So the unperturbed Hamiltonian is given by
\begin{gather}
H_0 = H_L +H_R +H_d \nonumber
\end{gather}
which is quadratic. The interacting term is in $H'$,
\begin{gather}
H' = H_T + H_{sd} \nonumber
\end{gather}
From now on we will use the following notation to label the trio-index $v_i =q_i n_i n'_i$. From equation of motion, we find the spin current from the right lead\cite{J-Ren}
\begin{gather}
I_{sR} = i\sum_{v } J_v[ \langle  d^{\dagger}_{n\uparrow} d_{n'\downarrow}a^{\dagger}_{q}\rangle -\langle a_{q} d^{\dagger}_{n'\downarrow} d_{n\uparrow}\rangle] \label{spin1}
\end{gather}
Now we define the Green's function on the Keldysh contour\cite{jauho}
\begin{gather}
G_{d,R}(\tau,\tau') = -i\sum_{v } J_v \langle T_c S d^{\dagger}_{n\uparrow}(\tau) d_{n'\downarrow}(\tau)a^{\dagger}_{q}(\tau') \rangle \label{lesser1}
\end{gather}
and
\begin{gather}
G_{R,d}(\tau,\tau') = -i\sum_{v } J_v \langle T_c S a_{q}(\tau)d^{\dagger}_{n'\downarrow}(\tau') d_{n\uparrow}(\tau')  \rangle \label{lesser2}
\end{gather}
so that
\begin{eqnarray}
I_{sR}(t) &=& - G_{d,R}^<(t,t)+G_{R,d}^<(t,t)\nonumber \\
&=& -i\sum_{v } J_v \langle d^{\dagger}_{n\uparrow}(t) d_{n'\downarrow}(t)a^{\dagger}_{q}(t) \rangle +h.c.\label{lesser}
\end{eqnarray}
where $G_{d,R}(\tau,\tau')$ is the Green's function connecting the right lead and the central scattering region. Here $T_c$ is the time ordering operator and $S$ is the S-matrix defined as
\begin{gather}
S = \exp(-i\int_c d\tau H'(\tau)) \nonumber
\end{gather}
where $H'$ is the interacting coupling term defined above. $\langle O \rangle$ is the expectation value of $O$ over ground state of $H_0$ which is non-interacting.

The Green's functions $G_{d,R}(\tau,\tau')$ and $G_{R,d}(\tau,\tau')$ are calculated in Appendix B. Using Eq.~(\ref{is}) and after making analytic continuation, we obtain
\begin{eqnarray}
I_{sR}(t)  &=&  \int dt'{\rm Tr}[G_{\uparrow}^r(t,t') {\bar \Sigma}_{R\uparrow}^<(t',t)\nonumber \\
&+&G_{\uparrow}^<(t,t') {\bar \Sigma}_{R\uparrow}^a(t',t)]+ h.c.  \label{Iright}
\end{eqnarray}
where ${\bar \Sigma}^r_{R\sigma} =i G_{{\bar \sigma}}^r .{\tilde \Sigma}_{R}^< + iG_{{\bar \sigma}}^< .{\tilde \Sigma}_{R}^a $, ${\bar \Sigma}^<_{R\sigma}=i G^<_{{\bar \sigma}}.{\tilde \Sigma}_{R}^>$, matrix $\tilde A$ is the transpose of $A$ in time domain,  and the Hadamard matrix product is used.

Notice that $G^<_{R,d}(t,t)=-[G^<_{d,R}(t,t)]^\dagger$, from Eq.~(\ref{a2}) in Appendix B we obtain another expression for $I_{sR}$,
\begin{eqnarray}
I_{sR}(t)  &=& -\int dt'{\rm Tr}[G_{\downarrow}^r(t,t') {\bar \Sigma}_{R\downarrow}^<(t',t)\nonumber \\
&+&G_{\downarrow}^<(t,t') {\bar \Sigma}_{R\downarrow}^a(t',t)]+ h.c.  \label{Iright1}
\end{eqnarray}

\bigskip
Now we calculate the spin current from the left lead. Defining spin density operator $N_{s}= \sum_k(c^{\dagger}_{k\uparrow} c_{k\uparrow} - c^{\dagger}_{k\downarrow} c_{k\downarrow})$, the equation of motion gives,
\begin{eqnarray}
I_{sL} = (1/2) \partial_t N_s = (1/2)(I_{\uparrow}-I_{\downarrow}) \nonumber
\end{eqnarray}
where in time domain
\begin{eqnarray}
I_\sigma(\tau) = \int d\tau'{\rm Tr} [G_{\sigma}(\tau,\tau')\Sigma_{L\sigma}(\tau',\tau)+h.c.] \nonumber
\end{eqnarray}
where $G_{\sigma}$ is defined in Eq.~(\ref{Gup-def}) in Appendix B. The self-energy of the left lead is defined as
\begin{eqnarray}
\Sigma_{L\uparrow mn} = \sum_k t_{k\uparrow n}t^*_{k\uparrow m} g_{L\uparrow}
\end{eqnarray}
where $g_{L\uparrow}$ is the Green's function of the left lead. After analytic continuation, we obtain
\begin{eqnarray}
I_\sigma(t) = {\rm Tr} [G_{\sigma}^r\Sigma^<_{L\sigma} +G_{\sigma}^< \Sigma_{L\sigma}^a +h.c.]_{tt} \label{spin-left}
\end{eqnarray}
where $G^r_{\sigma}$ and $G^<_{\sigma}$ are given by Eqs.~(\ref{Gr}) and (\ref{Gless}) in Appendix B.

Now we consider DC case. Since there is no charge current in the left lead, we should have $I_\uparrow= -I_\downarrow$ and $I_{sL}=I_\uparrow$. So we only need to calculate $I_\uparrow$. From Eq.~(\ref{spin-left}), the left spin current in energy domain is given by
\begin{eqnarray}
I_{sL} &&= \int \frac{dE}{2\pi} {\rm Tr} [(G_{\uparrow}^r(E)-G_{\uparrow}^a(E))\Sigma^<_{L\uparrow}(E) \nonumber \\
&&+G_{\uparrow}^<(E) (\Sigma_{L\uparrow}^a(E) -\Sigma_{L\uparrow}^r(E))]~. \label{up1}
\end{eqnarray}
The DC spin current from the right lead is given by
\begin{equation}
I_{sR}= \int \frac{dE}{2\pi} {\rm Tr}[G_{\uparrow}^r(E) {\bar \Sigma}_{R\uparrow}^<(E)
+G_{\uparrow}^<(E) {\bar \Sigma}_{R\uparrow}^a(E)+h.c.]
\label{IR2}
\end{equation}

From Eqs.~(\ref{up1}) and (\ref{IR2}), total spin current is found to be
\begin{eqnarray}
I_{s} &&= \int \frac{dE}{2\pi} {\rm Tr} [(G_{\uparrow}^r(E)-G_{\uparrow}^a(E))\Sigma^<_{\uparrow}(E) \nonumber \\
&&+G_{\uparrow}^<(E) (\Sigma_{\uparrow}^a(E) -\Sigma_{\uparrow}^r(E))]~ =0
\end{eqnarray}
where we have defined $\Sigma^{<,r}_{\uparrow}(E)= \Sigma^{<,r}_{L\uparrow}(E)+{\bar \Sigma}^{<,r}_{R\uparrow}(E)$. Hence the spin current is conserved as expected and $I_{sL} = -I_{sR}$.

From Eqs.~(\ref{Iright1}) and (\ref{spin-left}), it is easy to see that $I_{sR}-I_{\downarrow}=0$ which proves that $I_\uparrow= -I_\downarrow$ is indeed correct.

\bigskip
\section{Calculation of $G_{d,R}$}\label{GdR}
In this section, we work in the time domain. We first compute $G_{d,R}$ in the absence of left lead.
Expanding Eq.~(\ref{lesser1}) to the second order in ${J}_v$, we have
\begin{eqnarray}
&&G_{d,R}^{(2)}=-\sum_{v v_1} J_{v} \langle T_c d^{\dagger}_{n\uparrow}(\tau) d_{n'\downarrow}(\tau) a^\dagger_q(\tau)
\nonumber \\
&&\times \int d\tau_1 [J_{v_1} a_{q_1}(\tau_1) d^{\dagger}_{n_1'\downarrow}(\tau_1) d_{n_1\uparrow}(\tau_1) +h.c.]\rangle \nonumber \\
&&=\langle T_c \int d\tau_1\sum_{v v_1} J_{v} J_{v_1} d_{n'\downarrow}(\tau) d^{\dagger}_{n'_1\downarrow}(\tau_1) \nonumber \\
&& \times a_{q_1}(\tau_1) a^\dagger_{q}(\tau) d_{n_1\uparrow}(\tau_1) d^{\dagger}_{n\uparrow}(\tau) \rangle \nonumber \\
&&= i^2\int d\tau_1 \sum_{nn_1}{\bar\Sigma}_{0\uparrow nn_1}(\tau_1,\tau) g_{d n_1 n\uparrow}(\tau_1,\tau) \nonumber \\
&&=-{\rm Tr}[{\bar\Sigma}_{0\uparrow} g_{d\uparrow}]_{\tau \tau}\label{a1}
\end{eqnarray}
where
\begin{eqnarray}
 \Sigma_{Rn n_1 n' n_1'}(\tau,\tau') = \sum_{q q_1} J_{v_1} J_{v} g_{Rq q_1}(\tau,\tau')\label{self-r}
\end{eqnarray}
is the self-energy of the right lead, $g_{d\sigma}=1/(E-H_{d\sigma})$ is the bare Green's function of the scattering region with $H_{d\sigma}= \sum_{n} \epsilon_{n\sigma} d^{\dagger}_{n\sigma} d_{n\sigma}$, and $g_R$ is the Green's function of the right lead. Here
\begin{eqnarray}
{\bar \Sigma}_{0\uparrow n_1n}(\tau,\tau') = i\sum_{n_1' n'} g_{d n_1' n'\downarrow}(\tau,\tau') \Sigma_{Rn n_1 n' n_1'}(\tau',\tau)\nonumber
\end{eqnarray}
\begin{figure}
\includegraphics[width=8.5cm ]{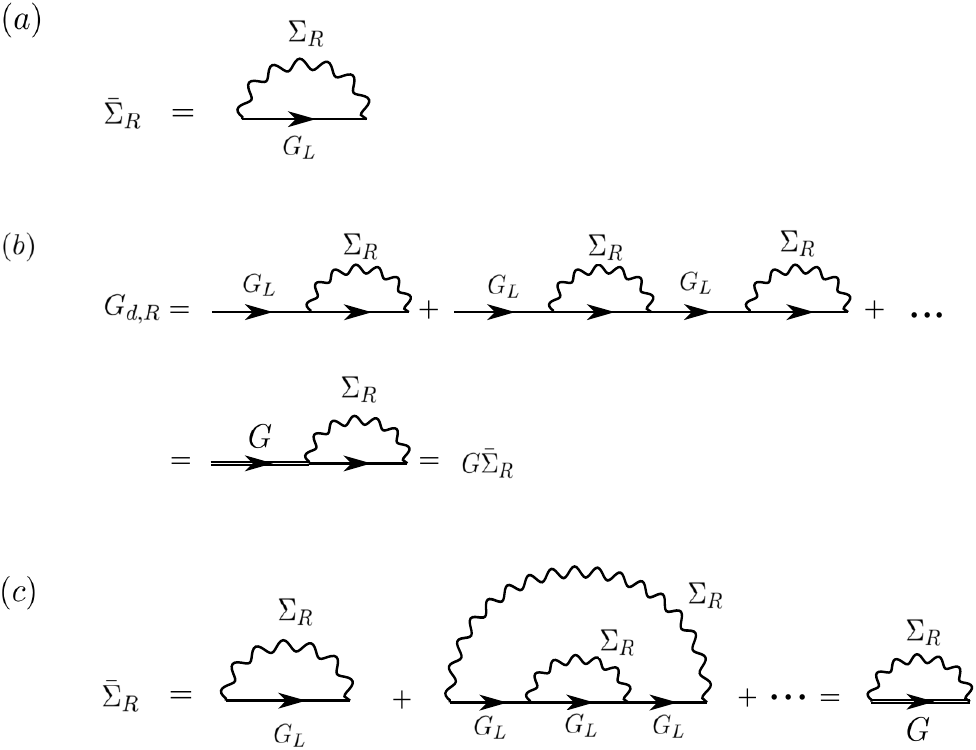}
\caption{(Color online) (a) Self-energy due to the first order Born approximation. (b) Calculation of $G_{d,R}$. (c) self-consistent Born approximation. }\label{figsm1}
\end{figure}

or
\begin{eqnarray}
{\bar \Sigma}_{0\uparrow}(\tau,\tau') = i g_{d\downarrow }(\tau,\tau') \Sigma_{R}(\tau',\tau)\nonumber
\end{eqnarray}
and the convention $v_i =q_i n_i n'_i$ was used.
Now we consider the fourth order of $J_v$ in Eq.~(\ref{lesser1}) (the integration for internal time index is implied),
\begin{eqnarray}
G_{d,R}^{(4)}&&=\sum_{v} J_{v} \langle T_c d^{\dagger}_{n\uparrow}(\tau) d_{n'\downarrow}(\tau) a^\dagger_q(\tau)
\nonumber \\
&&\times \sum_{v_1} [J_{v_1} a_{q_1}(\tau_1) d^{\dagger}_{n'_1\downarrow}(\tau_1) d_{n_1\uparrow}(\tau_1) +h.c.] \nonumber \\
&&\times \sum_{v_2}[J_{v_2} a_{q_2}(\tau_2) d^{\dagger}_{n'_2\downarrow}(\tau_2) d_{n_2\uparrow}(\tau_2) +h.c.] \nonumber\\
&&\times \sum_{v_3}[J_{v_3} a_{q_3}(\tau_3) d^{\dagger}_{n'_3\downarrow}(\tau_3) d_{n_3\uparrow}(\tau_3) +h.c.] \rangle \nonumber
\end{eqnarray}
where a typical pairing is given by
\begin{eqnarray}
 G_{d,R}^{(4)}&&=\sum_{v v_1 v_2 v_3} \langle T_c J_{v} J_{v_1}J_{v_2}J_{v_3} d_{n'\downarrow}(\tau) d^{\dagger}_{n_1'\downarrow}(\tau_1) \nonumber \\
&& \times a_{q}(\tau_1) a^\dagger_{q}(\tau)  d_{n_1\uparrow}(\tau_1) d^{\dagger}_{n_2\uparrow}(\tau_2) a_{q_3}(\tau_3)a^\dagger_{q_2}(\tau_2) \nonumber \\
&&\times  d_{n_2'\downarrow}(\tau_2) d^{\dagger}_{n_3'\downarrow}(\tau_3) d_{n_3\uparrow}(\tau_3) d^{\dagger}_{n\uparrow}(\tau) \nonumber \\
&&=-\sum_{n n_1 n_2 n_3} {\bar \Sigma}_{0\uparrow n_1 n}(\tau,\tau_1) g_{d n_1 n_2\uparrow}(\tau_1,\tau_2)\nonumber \\
&&\times  {\bar \Sigma}_{0\uparrow n_3 n_2}(\tau_3,\tau_2) g_{d n_3 n\uparrow}(\tau_3,\tau)\nonumber \\
&& =- {\rm Tr}[{\bar \Sigma}_{0\uparrow} g_{d\uparrow}{\bar \Sigma}_{0\uparrow} g_{d\uparrow}]_{\tau\tau} \nonumber
\end{eqnarray}
This pairing corresponds to first order Born approximation (Fig.~\ref{figsm1}(a)). Another pairing is shown in Fig.~\ref{figsm1}(c) which corresponds to self-consistent Born approximation (SCBA).
%Of course, we have to consider different combinations of pairing and the factor of $1/3!$.
When we consider the left lead, we replace $g_d$ by $G_{L}$, we find within first order Born approximation
\begin{eqnarray}
G_{d,R}&=& -{\rm Tr}[{\bar \Sigma}_{R\uparrow} G_{L\uparrow}]_{\tau \tau}
- {\rm Tr}[{\bar \Sigma}_{R\uparrow} G_{L\uparrow}{\bar \Sigma}_{R\uparrow} G_{L\uparrow}]_{\tau\tau}+... \nonumber \\
&=& -{\rm Tr} [{\bar \Sigma}_{R\uparrow} G_{\uparrow}]_{\tau \tau} \label{is}
\end{eqnarray}
where
\begin{eqnarray}
G_{L\uparrow} = g_{d\uparrow} +g_{d\uparrow}\Sigma_{L\uparrow}G_{L\uparrow} \label{GLup}
\end{eqnarray}
and
\begin{eqnarray}
G_{\uparrow} = g_{d\uparrow} +g_{d\uparrow}(\Sigma_{L\uparrow}+{\bar \Sigma}_{R\uparrow})G_{\uparrow} \label{Gup}
\end{eqnarray}
and
\begin{equation}
{\bar \Sigma}_{R\uparrow n_1n}(\tau,\tau') = i\sum_{n_1' n'} G_{L n_1' n'\downarrow}(\tau,\tau') \Sigma_{Rn n_1 n' n_1'}(\tau',\tau)\label{self-bar}
\end{equation}
To go beyond Born approximation, we have SCBA where Eq.~(\ref{is}) remains the same but Eq.~(\ref{self-bar}) becomes
\begin{equation}
{\bar \Sigma}_{R\uparrow n_1n}(\tau,\tau') = i\sum_{n_1' n'} G_{ n_1' n'\downarrow}(\tau,\tau') \Sigma_{Rn n_1 n' n_1'}(\tau',\tau)\label{self-bar1}
\end{equation}

Now we show that $G_{\uparrow}$ given by Eq.~(\ref{Gup}) can also be calculated from
\begin{eqnarray}
[G_{\uparrow}]_{nm} = -i\langle T_c S d_{n\uparrow} d^\dagger_{m\uparrow}\rangle \label{Gup-def}
\end{eqnarray}
using S-matrix expansion.

In the absence of the left lead, the second order term in $J_v$ in $G_{\uparrow}$ is
\begin{eqnarray}
&&\langle T_c d_{n\uparrow}(\tau) d^{\dagger}_{m\uparrow}(\tau') \sum_{v_1 }[J_{v_1} a_{q_1}(\tau_1) d^{\dagger}_{n_1'\downarrow}(\tau_1) d_{n_1\uparrow}(\tau_1) \nonumber \\
&&+h.c.] \sum_{v_2 }[J_{v_2} a_{q_2}(\tau_2) d^{\dagger}_{n_2'\downarrow}(\tau_2) d_{n_2\uparrow}(\tau_2) +h.c.]\rangle \nonumber
\end{eqnarray}
After taking care of the sign, we find
\begin{eqnarray}
G_{\uparrow} = g_{d\uparrow} + g_{d\uparrow} {\bar \Sigma_{R\uparrow}} g_{d\uparrow} +... \nonumber
\end{eqnarray}
When both leads are present, we have
\begin{eqnarray}
G_{\uparrow}=G_{L\uparrow} + G_{L\uparrow} {\bar \Sigma_{R\uparrow}} G_{\uparrow}~.\nonumber
\end{eqnarray}
Similarly, we find
\begin{eqnarray}
G_{\downarrow}=G_{L\downarrow}+ G_{L\downarrow} {\bar \Sigma}_{R\downarrow} G_{\downarrow} \nonumber
\end{eqnarray}
where ${\bar \Sigma}_{R\sigma}(\tau,\tau') =  iG_{{\bar\sigma}}(\tau,\tau'){ \Sigma}_{R}(\tau',\tau)$, $\sigma = (\uparrow, \downarrow)$ and ${\bar \sigma} = (\downarrow,\uparrow)$. After analytic continuation, Eq.~(\ref{GLup}) becomes
\begin{eqnarray}
G^r_{L\uparrow} = g^r_{d\uparrow} +g^r_{d\uparrow}\Sigma^r_{L\uparrow}G^r_{L\uparrow} \label{GLr}
\end{eqnarray}
and
\begin{eqnarray}
G^<_{L\uparrow} = G^r_{L\uparrow}\Sigma^<_{L\uparrow}G^a_{L\uparrow} \label{GLless}
\end{eqnarray}
Similarly, Eq.~(\ref{Gup}) becomes
\begin{eqnarray}
G^r_{\uparrow} = g^r_{d\uparrow} +g^r_{d\uparrow}(\Sigma^r_{L\uparrow}+{\bar \Sigma^r_{R\uparrow}})G^r_{\uparrow} \label{Gr}
\end{eqnarray}
and
\begin{eqnarray}
G^<_{\uparrow} = G^r_{\uparrow}(\Sigma^<_{L\uparrow}+{\bar \Sigma^<_{R\uparrow}})G^a_{\uparrow} \label{Gless}
\end{eqnarray}

Now we consider $G^<_{R,d}$. From the derivation of Eq.~(\ref{a1}) it is easy to see that to the second order in $J_v$,
\begin{eqnarray}
G_{R,d}=-{\rm Tr}[{\bar\Sigma}_{0\downarrow} g_{d\downarrow}]_{\tau \tau}
\end{eqnarray}
where
\begin{eqnarray}
{\bar \Sigma}_{0\downarrow}(\tau,\tau') = i g_{d\uparrow }(\tau,\tau') \Sigma_{R}(\tau',\tau)\nonumber
\end{eqnarray}
Including all orders of $J_v$ within SCBA, we find
\begin{eqnarray}
 G_{R,d}=-{\rm Tr} [{\bar \Sigma}_{R\downarrow} G_{\downarrow}]_{\tau \tau} \label{a2}
\end{eqnarray}

From Eqs.~(\ref{conti5}) and (\ref{conti6}), Eq.~(\ref{self-bar1}) becomes ${\bar \Sigma}^r_{R\sigma} =i G_{{\bar \sigma}}^r .{\tilde \Sigma}_{R}^< + iG_{{\bar \sigma}}^< .{\tilde \Sigma}_{R}^a $ and ${\bar \Sigma}^<_{R\sigma}=i G^<_{{\bar \sigma}}.{\tilde \Sigma}_{R}^>$.

\bigskip

%\noindent{\bf DC spin current from the right lead in SCBA}

The DC spin current from the right lead is given by Eq.~(3)
\begin{eqnarray}
I_{sR}&=& -\int dE {\rm Tr}[G_{\uparrow}^r(E) \Gamma_{L\uparrow}(E) G_{\uparrow}^a(E) (i{\bar \Sigma}_{R\uparrow}^<(E)\nonumber \\
&~& + 2f_{L\uparrow}(E){\rm Im}{\bar \Sigma}_{R\uparrow}^a(E))] . \label{isr2}
\end{eqnarray}
In SCBA the self-energies are determined self-consistently through following two equations,
\begin{eqnarray}
{\bar \Sigma}^<_{R\uparrow}(E)
&=& \int d\omega ~ G^r_{\downarrow}({\bar E})[i \Gamma_{L\downarrow}({\bar E}) f_{L\downarrow}({\bar E}) + {\bar \Sigma}^<_{R\uparrow}({\bar E})]\nonumber \\
&\times& G^a_{\downarrow}({\bar E})(1+f_R^B(\omega)) \Gamma_R(\omega)\nonumber
\end{eqnarray}
and
\begin{eqnarray}
{\bar \Sigma}^r_{R\uparrow}(E)&=&
\int d\omega G^r_{\downarrow}({\bar E})[f^B_R(\omega)
+  (-\Gamma_{L\downarrow}({\bar E}) f_{L\downarrow}({\bar E})\nonumber\\
&+& i{\bar \Sigma}^<_{R\uparrow}({\bar E}))G^a_{\downarrow}({\bar E})(i/2)]\Gamma_R(\omega)\nonumber
\end{eqnarray}
Note that BA has neglected ${\bar \Sigma}^<_{R\uparrow}({\bar E})$ in right hand side of the above two equations and ${\bar \Sigma}^r_R$ in $G^r_{\downarrow}$.

\section{Analytic Continuation}

We list here all the analytic continuations used in this work.  For $C = AB$ (matrix multiplication), we have\cite{jauho}
\begin{gather}
C^< = A^r B^< + A^< B^a, ~~ {\rm and } ~~ C^> = A^r B^> + A^> B^a \label{conti1}
\end{gather}
and
\begin{gather}
C^r = A^r B^r , ~~ {\rm and } ~~ C^a =  A^a B^a \label{conti2}
\end{gather}
For $C(\tau,\tau') = A(\tau,\tau') B(\tau,\tau')$ or $C=A.B$ (the Hadamard matrix product), we have\cite{jauho}
\begin{gather}
C^< = A^<. B^< , ~~ {\rm and } ~~ C^> = A^> .B^> \label{conti3}
\end{gather}
and
\begin{eqnarray}
&&C^r = A^r .B^< + A^< .B^r + A^r .B^r, \nonumber \\
&& C^a = A^a .B^< + A^< .B^a + A^a. B^a \label{conti4}
\end{eqnarray}
For $C(\tau,\tau') = A(\tau,\tau') B(\tau',\tau)$ or $C = A.{\tilde B}$ where ${\tilde B}(t_1,t_2) \equiv B(t_2,t_1)$, we have\cite{jauho}
\begin{gather}
C^< = A^< .{\tilde B}^> , ~~ {\rm and } ~~ C^> = A^> .{\tilde B}^< \label{conti5}
\end{gather}
\begin{eqnarray}
&& C^r = A^< .{\tilde B}^a + A^r .{\tilde B}^< , \nonumber \\
&& C^a = A^< .{\tilde B}^r + A^a .{\tilde B}^<  \label{conti6}
\end{eqnarray}
From Eqs.~(\ref{conti5}) and (\ref{conti6}), one can easily check the relation $C^> - C^< = C^r - C^a$ which must be satisfied.


\begin{thebibliography}{00}
\bibitem{note21} To be more precise, this DOS is defined as the local density of state for electron coming from a particular lead\cite{but1} and is also called injectivity.
\bibitem{note4} In obtaining Hamiltonian $H_R$, we start from the Heisenberg model and make Holstein-Primakoff transformation\cite{HP}. Hence there is an extra factor $\sqrt{2S_0}$ associated with $a_q$ and $a^\dagger_q$ in $H_{sd}$ which is dropped for the moment. Here $S_0$ is the length of the lattice spin.
\bibitem{note5} From now on, we absorb $1/(2\pi)^2$ from $dE$ and $d\omega$ into the unit of spin current. To resume the correct unit, the factor of $S_0 /(4\pi^2 \hbar) $ should be multiplied in calculating spin current.
\bibitem{note25} The spin current expression for SCBA can be found in Appendix A.
\bibitem{note2} We have used the relation $G^r - G^a = -i G^r \Gamma G^a$ which means that ${\rm Im} G^r <0$.
\bibitem{note7} We have used the identity $(f_{L\uparrow}(E)-1)f_{L\downarrow}({\bar E})= -(f_{L\uparrow}(E)-f_{L\downarrow}({\bar E}))f^B_L(\omega)$ where $f_{L\sigma}(\epsilon) = 1/[\exp(\beta_L (\epsilon-\mu_\sigma))+1]$, $f^B_L(\omega)= 1/[\exp(\beta_L (\omega+\mu_s))-1]$ is the effective Bose distribution of the left lead and $\mu_s= \mu_\uparrow-\mu_\downarrow$.
\bibitem{note22} Usually, the term "dynamical" is reserved for the time dependent transport. Here we use "dynamical" electron-magnon coupling to distinguish it from the static electron-magnon coupling which depends only on parameters of the closed system.
\bibitem{note10} We note that the work function of Pt is almost the largest among metals and semiconductors. Therefore insertion of intermediate layer of metals or semiconductors amounts to putting a potential well.
\bibitem{note13} The reduction of conductivity mismatch can be estimated by comparing $g^{-1}_{\rm Pt/YIG}$ with $g^{-1}_{\rm Pt/X} + g^{-1}_{\rm X/YIG} + G^{-1}_{\rm X}$ where $g_{\rm Pt/X}$ and $g_{\rm X/YIG}$ are spin mixing conductances of Pt/X and X/YIG interface, respectively, and $G_{\rm X}$ is the conductance of $X$ layer. It is clearly a sequential picture in quantum transport.
\bibitem{note11} Inserting a different material at the Pt/YIG interface leads to a change in $\alpha$ which we have neglected.
\bibitem{note14} This corresponds to a negative $v$ in Fig.~3(a) which suppresses the spin conductance.
\bibitem{note26} Since $E = (n\pi/a)^2 + k_n^2$, different subbands have different longitudinal energies.
\end{thebibliography}
\end{document}